\documentclass[preprint]{aastex}
\usepackage{chrisfigcaption}
\usepackage{chrisshortcuts}
\usepackage{graphicx}

\slugcomment{To be submitted to ApJS}

\shorttitle{AST/RO: HII regions}
\shortauthors{Walsh et al.}

\begin{document}

\title{An AST/RO Survey of CO(4-3) in ultracompact HII regions}

\author{Wilfred M. Walsh$^1$}

\affil{$^1$ Harvard-Smithsonian Center for Astrophysics, 60 Garden St.,
  MS-12, Cambridge, MA 02138}

\email{wwalsh@cfa.harvard.edu}

\begin{abstract}
The Antarctic Submillimeter Telescope and Remote Observatory (AST/RO)
has been used to observe 78 of the IRAS point sources identified by
Bronfman et al. (1996) as likely ultracompact HII regions. Results for
the \co{4} line at 461.041\ghz{} are presented. The 74
sources detected are bright and in many cases compact, making them
potentially suitable as pointing calibrators for single dish
submillimeter telescopes.
\end{abstract}

%\keywords{ISM:atoms --- ISM:general --- ISM:molecules \\[0.5cm]}

\section{Introduction}\label{s:intro}

Most star formation (SF) occurs within the giant molecular cloud
phase, and of particular interest is the behavior of the fragmented
interiors of these clouds in the stages immediately before, during and
after collapse into stellar objects. Both gaseous and dust components
play vital roles in all SF models, and this paper presents
submillimeter line observations of warm and dense gas near
ultracompact HII regions, the ionized gas surrounding early-type
stars. To identify the youngest stellar objects, far infrared (FIR)
observations from the IRAS satellite's near all-sky survey have been
used \citep{Wood1989ApJS...69..831W} and follow-up line studies have
successfully detected emission in many (sub)millimeter lines
(e.g. \citealt{Bronfman1996A&AS..115...81B, Snell2000ApJ...539L.101S,
Hatchell1998A&AS..133...29H}). Radio and IR continuum emission
originates from ultracompact HII regions, while submillimeter line
emission may come from hot molecular cores often found adjacent to one
or more ultracompact HII regions. These objects are considered the
best current tracer of ongoing SF.

An apparent problem with the understanding of ultracompact HII regions
is that their dynamical ages appear to be so short ($\sim10^4$\,yr) that
their numbers overpredict the current rate of SF. However the expected
lifetimes depend critically on the local distribution of gas and dust
(cf. \cite{Hollenbach1994ApJ...428..654H, dePree1995RMxAA..31...39D}),
of which little is known. Recent interferometric observations
\citep{Walsh1998MNRAS.301..640W, Koo1996ApJ...456..662K,
Kurtz1999ApJ...514..232K} suggest that at least some of the so-called
ultracompact HII regions may in fact have larger sizes and predicted
ages. Thus a better understanding of the distribution and dynamics of
molecular gas and dust around ultracompact HII regions is required. In
this paper we present results of a survey of 78 hot cores with the
Antarctic Submillimeter Telescope and Remote Observatory. This
survey is toward southern (Dec. $\lsim -20$\arcdeg\ so as to be
observable with AST/RO) sources in the list of
\citet{Bronfman1996A&AS..115...81B} with CS($2\rightarrow1$) emission
and associated IRAS luminosity and colours typical of compact HII
regions. The sources were further selected to have line profiles
indicative of inward or outward motions (e.g.,
\citealt{Mardones1998PhDT.........6M}) or have extended line wings
which may indicate the presence of bipolar outflows. This database is
likely to be representative of the early stages of massive SF. The
results of this survey will be combined with observations of other
submillimeter lines and continuum that will be analyzed in more detail
in a subsequent paper but will also be of use to the several new
southern submillimeter radiotelescopes as potential pointing and
calibration sources. Published properties of bright, compact
submillimeter sources in the southern sky are extremely sparse.

The observations are described in \S\,\ref{s:obs} and in
\S\,\ref{s:results} the data set is presented in the form of a table
of spectral line parameters and as plots of the full spectra. Small
maps of the brightest sources are also shown.

\section{Observations}
\label{s:obs}

The observations were performed during the austral winter season of
2002 at the Antarctic Submillimeter Telescope and Remote Observatory
\citep[AST/RO;][]{2001:Stark.Bally.Balm.Bania}, located at
2847\units{m} altitude in Amundsen-Scott South Pole Station.  This
site has very low water vapor, high atmospheric stability and a thin
troposphere making it exceptionally good for submillimeter
observations \citep{1997:Chamb.Lane.Stark, 1998:Lane}. AST/RO is a
1.7\units{m} diameter, offset Gregorian telescope capable of observing
at wavelengths between 200\microns{} and 1.3\units{mm}
\citep{1997:Stark.Chamb.Cheng.Ingal}. The receiver used was a
dual-channel SIS waveguide receiver
\citep{1992:Walke.Kooi.Chan.Leduc,1997:Honin.Haas.Hottg.Jacob} for
simultaneous 461--492\ghz{} and 807\ghz{} observations, with
double-sideband noise temperatures of 320--390\units{K} and
1050--1190\units{K}, respectively. Telescope efficiency,
$\eta_{\ell}$, estimated using moon scans, skydips, and measurements
of the beam edge taper, was 81\% at 461--492\ghz{} and 71\% at
807\ghz{}. The 807\ghz{} data will be presented, along with
observations in several other bands, in a subsequent
paper. Atmosphere-corrected system temperatures ranged from 700 to
4000\units{K} at 461--492\ghz{}.

A beam switching mode was used, with emission-free reference positions
chosen at least $20\arcmin$ from regions of interest, to make a small
map of points surrounding each source. These maps were repeated as
often as required to achieve suitable signal--to--noise. Emission from
the \co{4} and \co{7} lines at 461.041\ghz{} and 806.652\ghz{},
(together with the \cit{1} and \cit{2} lines at 492.262\ghz{} and
809.342\ghz{}), was imaged over the 78 regions with a spacing of a
half-beamwidth or less. The beam sizes (FWHM) were $103$--$109\arcsec$
at 461--492\ghz{} and $58\arcsec$ at 807\ghz{}
\citep{2001:Stark.Bally.Balm.Bania}. 

Two acousto-optical spectrometers 
\citep[AOSs;][]{1989:Schie.Tolls.Winne} were used as backends. The
AOSs had 1.07~\mhz{} resolution and 0.75\ghz{} effective bandwidth,
resulting in velocity resolution of 0.65\kms{} at 461\ghz{} and
0.37\kms{} at 807\ghz{}.  The data were smoothed to a uniform velocity
resolution of 1\kms{}.  The high frequency observations were made with
the CO $J = 7 \rightarrow 6$ line in the lower sideband (LSB). Since
the intermediate frequency of the AST/RO system is 1.5\ghz{}, the
\cit{2} line appears in the upper sideband (USB) and is superposed on
the observed LSB spectrum. The local oscillator frequency was chosen
so that the nominal line centers appear separated by 100\kms{} in the
double-sideband spectra.

The standard chopper wheel calibration technique was employed,
implemented at AST/RO by way of regular (every few minutes)
observations of the sky and two blackbody loads of known temperature
\citep{2001:Stark.Bally.Balm.Bania}. Atmospheric transmission was monitored
by regular skydips, and known, bright sources were observed every few
hours to further check calibration and pointing. At periodic intervals
and after tuning, the receivers were manually calibrated against a
liquid-nitrogen-temperature load and the two blackbody loads at
ambient temperature and about 100\units{K}. The latter process also
corrects for the dark current of the AOS optical CCDs. The intensity
calibration errors became as large as $\pm15$\% during poor weather
periods.

Once taken, the data in this survey were reduced using the COMB data
reduction package.  After elimination of scans deemed faulty for
various instrumental or weather-related reasons ($\lsim 7\%$ of the
total dataset), linear baselines were removed from the spectra in all
species by excluding regions where the \cs{2} spectra of
\citet{Bronfman1996A&AS..115...81B} showed emission within twice the
FWHM of Gaussian fits to the \cs{2} line.  This allowed known emission
to be readily excluded from the baseline fitting procedure.

While the original intent was to make \trms{} as uniform as possible
across all source maps, this was not always possible.  For the \co{4}
transition, \trms{} in 1\kms{} wide channels with no spatial smoothing
is on average $\lsim 0.75 \units{K}$.

\section{Results}
\label{s:results}

AST/RO's pointing model \citep{2001:Stark.Bally.Balm.Bania} is
currently determined by observing a small number of sources for a
24~hr period so as to obtain full coverage of the sky in
azimuth. However these sources do not cover a wide range in
elevation. As the residual pointing uncertainty after the application
of the pointing model at AST/RO is between one beamwidth in the
frequency used to determine the pointing model and one arcminute, it
is a major aim of this work to identify a larger sample of compact
sources, distributed over the sky, that may be used for pointing
calibration. Therefore small images of a few square arcminutes were
made of the sample with half-beam spacing. Fig.~1 displays the
brightest spectrum observed in the vicinity of each source and Fig.~2
shows the resulting images.

Table~\ref{t:results} lists the results of Gaussian fits to the
observed lines shown in Fig.~1. The first column is a shortened name
based on the Galactic longitude, column two is the IRAS source name,
columns three and four are the equinox J2000 coordinates. Column five
is the peak antenna temperature as estimated by the Gaussian fit, with
an uncertainty due to the fitting. The actual intensity calibration
error of AST/RO is generally larger (\S\,\ref{s:obs}). Columns six
to eight are the central velocity, integrated line intensity (in
K\kms{}) and FWHM of the line as estimated by the Gaussian fitting,
respectively. Fig.~1 shows that many of the line profiles can be
approximately represented by a Gaussian form, leading to a reasonable
estimate of the line strength, width and central velocity. The
Gaussians, fit to a range of channels 10\kms{} either side of the FWHM
of the \citet{Bronfman1996A&AS..115...81B} Gaussian fit to the \cs{2}
line, clearly do not provide a precise model of the profiles, and
further analysis of the spectra should refer to the original data,
available online.

Fig.~1 shows that the great majority, 95\%, of the sources are
detected in the \co{4} line and 86\% of them have line strengths
brighter than 5~K. Thus the \co{4} line is a readily-detectable tracer
of molecular material around ultracompact HII regions, and may be used
as a kinematic tracer and for distance determination. The profiles are
in nearly all cases characterized by a single component, whose
$V_{\mathrm {lsr}}$ is in all but two cases the same as that of the
\cs{2} line, within the fitting uncertainties. Those sources not
detected at a level of several times the RMS noise per channel are
indicated in the table as upper limits, the level of which is
estimated to be 4.5 times the RMS per channel.

Images of a selection of the brightest sources detected in the \co{4}
line are shown in Fig.~2. The images were formed by simply summing the
emission in channels within the FWHM of the
\citet{Bronfman1996A&AS..115...81B} Gaussian fit to the \cs{2}
line. The images have been gridded onto a surface that over-samples the
observed pointing centers by a factor of three using a Gaussian
smoothing function with FWHM 30\% larger than the beam, and weighted
using cone interpolation with a similarly-sized interpolation
radius. Bright molecular emission in the vicinity of ultracompact HII
regions often originates from hot cores, which are expected to be
relatively point-like compared with the AST/RO beam, except in those
cases where an outflow is seen, where more than one molecular core is
present, or if the dense molecular region is genuinely
extended. Recent VLA results
\citep{Kurtz1999ApJ...514..232K} show some ultracompact HII regions to
lie within larger structures that may also contain extended molecular
material. From Fig.~2 it can be seen that at least 50\% of the sources
in the present sample are unresolved by AST/RO and can potentially be
used by single dish telescopes for pointing purposes. 
%It should be
%noted that a small number of the images that appear extended in Fig.~2
%are simply those with poor signal-to-noise and upon deeper
%integration prove also to be compact.

The point-like images shown in fig.~2 can be used to select sub-samples
for pointing purposes while extended structures may benefit from
future mapping. These data will be combined with and analyzed in the
light of measurements made in several other lines with a variety of
opacities in a future paper.

\begin{deluxetable}{llrrrrrr}
\tabletypesize{\footnotesize}
\tablewidth{0pt}
\tablecaption{Results of AST/RO observations}
\tablehead{
\colhead{Name}&\colhead{IRAS name}&\colhead{R.A. (J2000)}&\colhead{Dec. (J2000)}&\colhead{T$^{\rm A}_{*}$ (K)}&
\colhead{V$_{\rm lsr}$}&\colhead{$I$}&\colhead{$\Delta V$}}
\startdata
G268.522 &    9028-4837  &	08:59:29.82 & -48:13:17.31 &          2.8 $\pm$  0.6\tablenotemark{a} &   - &   - &   - \\
%	 &               &	            &	           & 	5.6  $\pm$   1.4 &   -0.2  $\pm$ 0.1  &  3.8   & -0.6	 \\ 
%	 &               &	            &	           & 	2.0  $\pm$   1.0 &   4.0   $\pm$ 0.5  &  3.8   & 0.6	 \\
G269.854 &    9094-4803  &	09:11:08.40 & -48:15:59.16 &          2.9 $\pm$  0.5 &   78.8 $\pm$  0.1 &   4.1 &   1.3 \\
%	 &               &	            &	           & 	5.8  $\pm$   1.1 &   75.2  $\pm$ 0.1  &  9.7   & 1.6	 \\
G281.586 &    10031-5632 &	10:04:56.23 & -56:46:36.85 &          7.3 $\pm$  0.1 &   -1.2 $\pm$  0.1 &  45.2 &   5.8 \\
%	 &               &	            &	           & 	6.2  $\pm$   0.6 &   0.9   $\pm$ 0.1  &  14.4  & 2.2	 \\
G285.259 &    10295-5746 &	10:31:28.36 & -58:02:07.47 &         23.3 $\pm$   0.1 &     4.8 $\pm$  0.0 & 129.9  & 5.2 \\
%	 &               &	            &	           & 	11.1 $\pm$   0.5 &   4.0   $\pm$ 0.1  &  38.4  & 3.3	 \\ 
%	 &               &	            &	           & 	31.6 $\pm$   5.2 &   4.3   $\pm$ 0.1  &  113.5 & 3.4	 \\
G291.274 &    11097-6102 &	11:11:52.66 & -61:18:35.38 &         25.3 $\pm$  0.2 &  -20.9 $\pm$  0.0 & 260.7 &   9.7 \\
%	 &               &	            &	           & 	10.5 $\pm$   0.2 &   -22.0 $\pm$ 0.0  &  40.7  & 3.7	 \\ 
%	 &               &	            &	           & 	22.5 $\pm$   0.5 &   -21.6 $\pm$ 0.1  &  147.1 & 6.2	 \\
G297.725 &    11590-6452 &	12:03:22.80 & -64:09:57.25 &          3.1 $\pm$  0.8 &   -9.2 $\pm$  0.1 &   2.3 &   0.7 \\
G301.116 &    12331-6134 &	12:36:02.05 & -61:51:05.41 &          9.6 $\pm$  0.2 &  -39.0 $\pm$  0.1 &  81.6 &   8.0 \\
%	 &               &	            &	           & 	11.7 $\pm$   1.2 &   -38.9 $\pm$ 0.1  &  32.2  & 2.6	 \\
G301.134 &    12326-6245 &	12:35:33.87 & -63:02:29.05 &         13.0 $\pm$  0.2 &  -37.6 $\pm$  0.1 &  87.0 &   6.3 \\
G301.722 &               &	12:41:17.43 & -61:44:10.45 &         10.9 $\pm$  0.2 &  -38.6 $\pm$  0.0 &  52.6 &   4.5 \\
%	 &               &	            &	           & 	15.8 $\pm$   0.9 &   -40.3 $\pm$ 0.1  &  47.5  & 2.8	 \\
G301.731 &    12383-6128 &	12:41:17.42 & -61:44:38.99 &          9.4 $\pm$  0.3 &  -37.9 $\pm$  0.1 &  53.9 &   5.4 \\
G305.194 &               &	13:11:13.10 & -62:44:56.10 &         11.7 $\pm$  0.2 &  -33.9 $\pm$  0.1 &  87.3 &   7.0 \\
%	 &               &	            &	           & 	11.4 $\pm$   2.0 &   -42.8 $\pm$ 0.1  &  16.3  & -1.3	 \\
G307.559 &    13080-6229 &	13:32:30.47 & -63:04:48.90 &          6.7 $\pm$  0.2 &  -34.8 $\pm$  0.1 &  28.2 &   3.9 \\
%	 &               &	            &	           & 	9.9  $\pm$   6.5 &   -33.8 $\pm$ 0.0  &  4.1   & -0.4	 \\ 
%	 &               &	            &	           & 	6.6  $\pm$   1.7 &   -43.6 $\pm$ 0.1  &  4.5   & 0.6	 \\
G307.560 &    13291-6249 &	13:32:30.69 & -63:05:17.36 &          7.3 $\pm$  0.3 &  -33.4 $\pm$  0.1 &  32.9 &   4.2 \\
%	 &               &	            &	           & 	3.4  $\pm$   1.1 &   -44.1 $\pm$ 0.1  &  2.2   & 0.6	 \\
G309.920 &    13471-6120 &	13:50:41.53 & -61:35:10.53 &          8.8 $\pm$  0.3 &  -55.6 $\pm$  0.1 &  31.2 &   3.3 \\
%	 &               &	            &	           & 	12.0 $\pm$   1.7 &   -56.6 $\pm$ 0.2  &  38.1  & 3.0	 \\
G310.142 &    13484-6100 &	13:51:57.63 & -61:15:45.78 &          7.9 $\pm$  0.3 &  -54.9 $\pm$  0.1 &  78.0 &   9.2 \\
G312.599 &               &	14:13:13.88 & -61:16:18.26 &          6.7 $\pm$  0.2 &  -62.6 $\pm$  0.1 &  47.0 &   6.6 \\
G312.596 &    14095-6102 &	14:13:13.61 & -61:16:47.12 &          6.2 $\pm$  0.3 &  -60.8 $\pm$  0.1 &  29.8 &   4.5 \\
%	 &               &	            &	           & 	11.7 $\pm$   1.4 &   -63.5 $\pm$ 0.1  &  18.1  & 1.5	 \\ 
%	 &               &	            &	           & 	1.9  $\pm$   0.8 &   -53.2 $\pm$ 0.7  &  6.6   & 3.2	 \\
G318.047 &    14498-5856 &	14:53:41.34 & -59:08:53.77 &          8.9 $\pm$  0.2 &  -48.5 $\pm$  0.1 &  65.7 &   7.0 \\
G319.163 &    14593-5852 &	15:03:13.25 & -59:03:53.96 &          6.9 $\pm$  0.1 &  -20.4 $\pm$  0.1 &  81.0 &  11.0 \\
%	 &               &	            &	           & 	11.2 $\pm$   2.0 &   -31.4 $\pm$ 0.1  &  8.3   & 0.7	 \\ 
%	 &               &	            &	           & 	0.7  $\pm$   0.3 &   -32.7 $\pm$ 6.9  &  14.1  & 19.2	 \\
G320.674 &    15068-5733 &	15:10:43.24 & -57:44:46.67 &          4.3 $\pm$  0.3 &  -57.3 $\pm$  0.1 &  19.4 &   4.3 \\
G321.719 &    15100-5613 &	15:13:49.44 & -56:24:55.02 &          7.9 $\pm$  0.2 &  -39.6 $\pm$  0.1 &  54.5 &   6.5 \\
G322.933 &    15165-5524 &      15:20:21.12 & -55:35:04.41 &          6.1 $\pm$  0.6 &  -38.9 $\pm$  0.1 &   6.6 &   1.0 \\
G324.201 &    15290-5546 &	15:32:53.62 & -55:56:12.40 &          8.9 $\pm$  0.2 &  -85.7 $\pm$  0.1 &  77.7 &   8.2 \\
G326.466 &               &	15:43:17.74 & -54:07:00.09 &          8.6 $\pm$  0.1 &  -40.4 $\pm$  0.1 & 108.6 &  11.8 \\
%	 &               &	            &	           & 	13.0 $\pm$   2.0 &   -42.1 $\pm$ 0.1  &  16.4  & 1.2	 \\ 
%	 &               &	            &	           & 	4.8  $\pm$   0.8 &   -40.6 $\pm$ 0.6  &  32.9  & 6.5	 \\
G326.474 &    15394-5358 &	15:43:17.83 & -54:07:32.62 &          7.5 $\pm$  0.2 &  -38.9 $\pm$  0.0 &  22.9 &   2.9 \\
G326.655 &    15408-5356 &	15:44:42.79 & -54:05:56.00 &         21.4 $\pm$  0.2 &  -38.1 $\pm$  0.0 & 224.8 &   9.9 \\
G328.306 &               &	15:54:06.03 & -53:11:07.59 &          2.0 $\pm$  0.6 &   -8.9 $\pm$  0.1 &   1.7 &   0.8 \\
%	 &               &	            &	           & 	15.6 $\pm$   10.4&   -13.1 $\pm$ 0.1  &  5.7   & -0.4	 \\ 
%	 &               &	            &	           & 	9.4  $\pm$   2.3 &   -10.8 $\pm$ 0.1  &  9.4   & 0.9	 \\
G328.307 &    15502-5302 &	15:54:06.01 & -53:11:36.51 &         12.9 $\pm$  0.2 &  -91.2 $\pm$  0.1 & 116.2 &   8.5 \\
G328.809 &    15520-5234 &	15:55:48.49 & -52:42:40.20 & 	3.4  $\pm$   0.6 &   -101.3$\pm$ 0.1  & 3.7    & 1.0	 \\ 
%	 &               &	            &	           & 	10.9 $\pm$   2.3 &   -116.0$\pm$ 0.1  & 13.4   & 1.2	 \\ 
%	 &               &	            &	           & 	11.3 $\pm$   2.8 &   -109.4$\pm$ 0.1  & 7.2    & -0.6	 \\
G329.337 &    15567-5236 &	16:00:32.89 & -52:44:47.59 &         13.7 $\pm$  0.0 &  -105.7 $\pm$  0.0 &  94.8 &   6.5 \\
G329.066 &    15573-5307 &	16:01:09.72 & -53:16:01.76 &          4.9 $\pm$  0.3 &  -47.3 $\pm$  0.1 &  14.6 &   2.8 \\
G329.404 &    15596-5301 &	16:03:31.25 & -53:09:26.83 &          6.2 $\pm$  0.3 &  -73.7 $\pm$  0.1 &  36.0 &   5.5 \\
%	 &               &	            &	           & 	6.0  $\pm$   0.5 &   -74.2 $\pm$ 0.1  &  9.8   & 1.5	 \\ 
G335.582 &    16272-4837 &	16:30:56.40 & -48:43:46.39 &          3.7 $\pm$  0.5 &  -66.4 $\pm$  0.1 &   7.2 &   1.8 \\
%	 &               &	            &	           & 	5.5  $\pm$   0.9 &   -57.8 $\pm$ 0.1  &  5.4   & 0.9	 \\ 
%	 &               &	            &	           & 	8.8  $\pm$   2.5 &   -65.3 $\pm$ 0.1  &  4.0   & 0.4	 \\
G330.883 &    16065-5158 &	16:10:21.83 & -52:06:01.98 &         17.2 $\pm$  0.2 &  -61.5 $\pm$  0.0 & 132.0 &   7.2 \\
G330.946 &    16060-5146 &	16:09:48.30 & -51:54:52.36 &          8.7 $\pm$  0.2 &  -88.3 $\pm$  0.1 &  74.0 &   8.0 \\
G331.126 &    16071-5142 &	16:10:56.83 & -51:50:24.06 &          6.3 $\pm$  0.2 &  -86.1 $\pm$  0.1 &  42.5 &   6.4 \\
%	 &               &	            &	           & 	13.5 $\pm$   2.3 &   -86.8 $\pm$ 0.2  &  29.5  & 2.1	 \\
G332.153 &               &	16:16:39.32 & -51:16:28.40 &         10.6 $\pm$  0.2 &  -54.9 $\pm$  0.1 &  99.9 &   8.8 \\
%	 &               &	            &	           & 	14.3 $\pm$   0.6 &   -59.9 $\pm$ 0.0  &  10.3  & -0.7	 \\ 
%	 &               &	            &	           & 	8.5  $\pm$   1.2 &   -53.9 $\pm$ 0.1  &  14.5  & 1.6	 \\
G332.293 &    16119-5048 &	16:15:45.15 & -50:56:02.83 &          9.9 $\pm$  0.3 &  -47.7 $\pm$  0.0 &  31.5 &   3.0 \\
%	 &               &	            &	           & 	12.2 $\pm$   1.9 &   -48.1 $\pm$ 0.1  &  14.5  & -1.1	 \\
G332.653 &    16158-5055 &	16:19:40.73 & -51:03:10.97 &         11.3 $\pm$  0.2 &  -47.4 $\pm$  0.1 & 103.9 &   8.6 \\
%	 &               &	            &	           & 	10.0 $\pm$   0.8 &   -48.0 $\pm$ 0.2  &  45.1  & 4.3	 \\ 
%	 &               &	            &	           & 	17.4 $\pm$   4.6 &   -51.3 $\pm$ 0.1  &  8.4   & -0.5	 \\
G332.831 &    16164-5046 &	16:20:14.28 & -50:53:19.87 &          8.8 $\pm$  0.2 &  -55.3 $\pm$  0.1 & 117.7 &  12.6 \\
%	 &               &	            &	           & 	12.3 $\pm$   1.0 &   -56.0 $\pm$ 0.2  &  67.9  & 5.2	 \\
G333.129 &    16172-5028 &	16:21:00.60 & -50:35:19.84 &         22.4 $\pm$  0.2 &  -50.5 $\pm$  0.0 & 224.7 &   9.4 \\
%	 &               &	            &	           & 	15.6 $\pm$   1.2 &   -49.5 $\pm$ 0.3  &  109.8 & 6.6	 \\
G333.306 &    16177-5018 &	16:21:30.61 & -50:25:04.40 &         21.6 $\pm$  0.2 &  -49.0 $\pm$  0.0 & 225.1 &   9.8 \\
%	 &               &	            &	           & 	15.7 $\pm$   3.7 &   -47.9 $\pm$ 0.1  &  14.3  & 0.9	 \\
G337.164 &    16351-4722 &	16:36:20.15 & -47:24:29.43 &          4.5 $\pm$  0.2 &  -65.0 $\pm$  0.2 &  55.1 &  11.6 \\
%	 &               &	            &	           & 	24.5 $\pm$   5.7 &   -72.5 $\pm$ 0.1  &  26.8  & 1.0	 \\ 
%	 &               &	            &	           & 	17.6 $\pm$   4.7 &   -60.8 $\pm$ 0.1  &  11.9  & 0.6	 \\
G337.703 &    16348-4654 &	16:38:33.29 & -47:01:20.00 &          4.9 $\pm$  0.3 &  -52.7 $\pm$  0.2 &  34.8 &   6.7 \\
G338.569 &    16385-4619 &	16:42:14.29 & -46:25:28.13 &         10.2 $\pm$  0.3 &  -114.6 $\pm$  0.1 &  54.0 &   5.0 \\
G339.622 &    16424-4531 &	16:46:06.65 & -45:36:49.42 &          5.7 $\pm$  0.2 &  -32.6 $\pm$  0.1 &  40.2 &   6.6 \\
G340.053 &    16445-4516 &	16:48:11.89 & -45:21:32.25 &          9.1 $\pm$  0.2 &  -51.0 $\pm$  0.1 &  77.2 &   8.0 \\
G340.248 &    16458-4512 &	16:49:30.26 & -45:17:49.66 &          8.3 $\pm$  0.2 &  -49.5 $\pm$  0.1 &  71.0 &   8.0 \\
%	 &               &	            &	           & 	17.7 $\pm$   1.5 &   -51.4 $\pm$ 0.2  &  79.9  & 4.2	 \\ 
%	 &               &	            &	           & 	18.5 $\pm$   4.9 &   -39.7 $\pm$ 0.1  &  12.2  & 0.6	 \\
G341.932 &    16510-4347 &	16:54:37.12 & -43:51:55.92 &          7.0 $\pm$  0.3 &  -42.0 $\pm$  0.1 &  33.4 &   4.5 \\
%	 &               &	            &	           & 	6.6  $\pm$   0.2 &   -41.2 $\pm$ 0.2  &  62.9  & 9.0	 \\ 
%	 &               &	            &	           & 	8.3  $\pm$   0.5 &   -42.0 $\pm$ 0.1  &  23.0  & 2.6	 \\
G342.697 &               &      16:56:04.02 & -43:04:13.54 &          6.1 $\pm$  0.4 &  -40.9 $\pm$  0.1 &  25.8 &   3.9 \\
G342.704 &    16524-4300 &	16:56:01.29 & -43:04:43.95 &          5.5 $\pm$  0.9 &  -41.2 $\pm$  0.1 &   4.3 &   0.7 \\
G343.126 &    16547-4247 &	16:58:16.90 & -42:51:37.00 &          8.4 $\pm$  0.5 &  -28.4 $\pm$  0.1 &  25.4 &   2.9 \\
%	 &               &	            &	           & 	10.7 $\pm$   0.9 &   -30.8 $\pm$ 0.1  &  19.5  & 1.7	 \\ 
%	 &               &	            &	           & 	8.9  $\pm$   3.1 &   -30.7 $\pm$ 0.1  &  5.0   & 0.5	 \\
G345.001 &    17016-4124 &	17:05:09.79 & -41:28:34.07 &          6.5 $\pm$  0.8 &  -85.7 $\pm$  0.1 &   9.5 &   1.4 \\
G345.208 &    16571-4029 &	17:00:35.41 & -40:33:31.17 &         23.2 $\pm$  0.3 &  -13.9 $\pm$  0.0 & 133.9 &   5.4 \\
%	 &               &	            &	           & 	10.7 $\pm$   1.4 &   -10.8 $\pm$ 0.2  &  29.2  & 2.6	 \\
%	 &               &	17:04:26.83 & -40:45:57.05 & 	13.3 $\pm$   0.1 &   -17.3 $\pm$ 0.0  &  82.5  & 5.8	 \\
G345.482 &               &	17:04:26.83 &-40:45:57.05  &         21.4 $\pm$  0.4 &  -17.2 $\pm$  0.1 & 155.5 &   6.8 \\
G345.490 &    17009-4042 &	17:04:29.50 & -40:46:25.47 &         13.2 $\pm$  0.3 &  -16.3 $\pm$  0.1 & 117.2 &   8.4 \\
%	 &               &	            &	           & 	6.1  $\pm$   0.5 &   -17.2 $\pm$ 0.1  &  22.3  & 3.4	 \\
G345.494 &    16562-3959 &	16:59:41.88 & -40:03:44.10 &         19.3 $\pm$  0.2 &  -10.8 $\pm$  0.1 & 192.6 &   9.4 \\
%	 &               &	            &	           & 	33.6 $\pm$   7.1 &   -13.5 $\pm$ 0.1  &  30.2  & 0.8	 \\
G345.499 &    17008-4040 &	17:04:20.41 & -40:44:25.77 &         13.1 $\pm$  0.3 &  -16.4 $\pm$  0.1 &  71.1 &   5.1 \\
%	 &               &	            &	           & 	5.5  $\pm$   0.6 &   -15.7 $\pm$ 0.2  &  19.3  & 3.3	 \\
%	 &               &	17:04:23.06 & -40:43:56.31 & 	15.0 $\pm$   0.3 &   -16.7 $\pm$ 0.1  &  114.6 & 7.2	 \\
G345.505 &    17008-4040 &	17:04:23.06 & -40:43:56.31 &         21.9 $\pm$  0.4 &  -17.1 $\pm$  0.1 & 172.8 &   7.4 \\
G345.717 &    16596-4012 &	17:03:06.30 & -40:17:08.73 &          7.1 $\pm$  0.6 &   -9.0 $\pm$  0.1 &  13.6 &   1.8 \\
%	 &               &	            &	           & 	8.0  $\pm$   1.9 &   -10.9 $\pm$ 0.1  &  5.2   & 0.6	 \\
G348.236 &    17149-3916 &	17:18:23.91 & -39:19:10.19 &         16.6 $\pm$  0.3 &  -10.8 $\pm$  0.1 & 109.7 &   6.2 \\
G348.534 &    17158-3901 &	17:19:16.17 & -39:04:26.09 &         10.3 $\pm$  0.2 &  -11.2 $\pm$  0.1 & 117.0 &  10.7 \\
G348.548 &               &	17:19:16.05 & -39:03:55.67 &         22.9 $\pm$  0.5 &  -11.0 $\pm$  0.1 & 208.2 &   8.5 \\
%	 &               &	17:19:16.05 & -39:03:55.67 & 	11.3 $\pm$   0.3 &   -11.7 $\pm$ 0.1  &  115.4 & 9.6	 \\ 
%	 &               &	            &	           & 	16.5 $\pm$   5.7 &   -17.9 $\pm$ 0.1  &  13.4  & -0.8	 \\
G350.103 &    17160-3707 &	17:19:26.32 & -37:10:54.75 &          9.1 $\pm$  0.2 &  -68.0 $\pm$  0.1 & 124.2 &  12.8 \\
G350.504 &    17136-3617 &	17:17:02.20 & -36:21:08.60 &          3.1 $\pm$  0.6 &   15.7 $\pm$  0.2 &   5.7 &   1.7 \\
%G350.504 &    17136-3617 &	17:17:02.20 & -36:21:08.60 &         21.9 $\pm$  0.4 &  -17.1 $\pm$  0.1 & 172.8 &   7.4 \\
G351.776 &    17233-3606 &	17:26:44.44 & -36:09:26.63 &          9.1 $\pm$  0.2 &   -1.4 $\pm$  0.1 & 112.9 &  11.6 \\
G352.630 &    17278-3541 &	17:31:13.88 & -35:44:09.07 &          4.9 $\pm$  0.2 &   -0.4 $\pm$  0.2 &  62.9 &  11.9 \\
G353.410 &               &	17:30:26.47 & -34:41:09.16 &         26.5 $\pm$  0.6 &  -15.0 $\pm$  0.1 & 313.6 &  11.1 \\
%	 &               &	            &	           & 	10.1 $\pm$   2.4 &   -20.6 $\pm$ 0.7  &  67.6  & 6.3	 \\
G353.416 &    17271-3439 &	17:30:28.87 & -34:41:40.53 &         11.8 $\pm$  0.3 &  -14.6 $\pm$  0.1 & 153.8 &  12.3 \\
%	 &               &	            &	           & 	11.3 $\pm$   0.6 &   -14.3 $\pm$ 0.2  &  80.9  & 6.7	 \\
G357.552 &    17385-3116 &	17:41:49.73 & -31:18:22.66 &         14.1 $\pm$  0.4 &    3.4 $\pm$  0.1 &  54.0 &   3.6 \\
G000.665 &    17441-2822 &	17:47:19.66 & -28:23:08.21 & 	4.5  $\pm$   1.0\tablenotemark{a} &   -    &  -   &  \\
G005.633 &    17545-2357 &	17:57:33.60 & -23:58:15.12 & 	4.8  $\pm$   0.9\tablenotemark{a} &   -  &  -   & -	 \\
G005.888 &    17574-2403 &	18:00:32.09 & -24:04:02.75 &         18.5 $\pm$  0.4 &   10.0 $\pm$  0.1 & 123.9 &   6.3 \\
G008.139 &    17599-2148 &	18:03:00.39 & -21:48:04.92 &          7.8 $\pm$  0.4 &   21.3 $\pm$  0.2 &  44.6 &   5.4 \\
G009.615 &    18032-2032 &	18:06:13.42 & -20:31:47.22 &          9.3 $\pm$  0.4 &    7.0 $\pm$  0.1 &  58.8 &   6.0 \\
G010.157 &    18064-2020 &	18:09:24.44 & -20:19:27.99 &          7.4 $\pm$  0.3 &    9.8 $\pm$  0.3 & 107.6 &  13.7 \\
G010.466 &    18056-1952 &	18:08:36.63 & -19:52:03.35 &          6.9 $\pm$  0.4 &   72.3 $\pm$  0.3 &  62.6 &   8.5 \\
G011.936 &    18110-1854 &	18:14:00.34 & -18:53:22.22 & 	5.0  $\pm$   1.8\tablenotemark{a} &  -  &  -   & -	 \\
\enddata
\label{t:results}
\tablenotetext{a}{Upper limit: see text.}
\end{deluxetable}

\acknowledgments
\label{s:ack}

I thank K. Brooks and G. Garay for suggesting the sample; K. Xiao,
C. Martin and A. Stark for help with the observations; C. Walker and
the receiver group at the U. of Arizona for their assistance;
R. Schieder, J. Stutzki, and colleagues at U. K\"{o}ln for their AOSs;
J. Kooi and R. Chamberlin of Caltech, G. Wright of PacketStorm
Communications, and K. Jacobs of U. K\"{o}ln for their work on the
instrumentation. This research was supported in part by the National
Science Foundation under a cooperative agreement with the Center for
Astrophysical Research in Antarctica (CARA), grant number NSF OPP
89-20223. CARA is a National Science Foundation Science and Technology
Center. Support was also provided by NSF grant number OPP-0126090.

\bibliography{hotcores}
\bibliographystyle{apj}

\clearpage

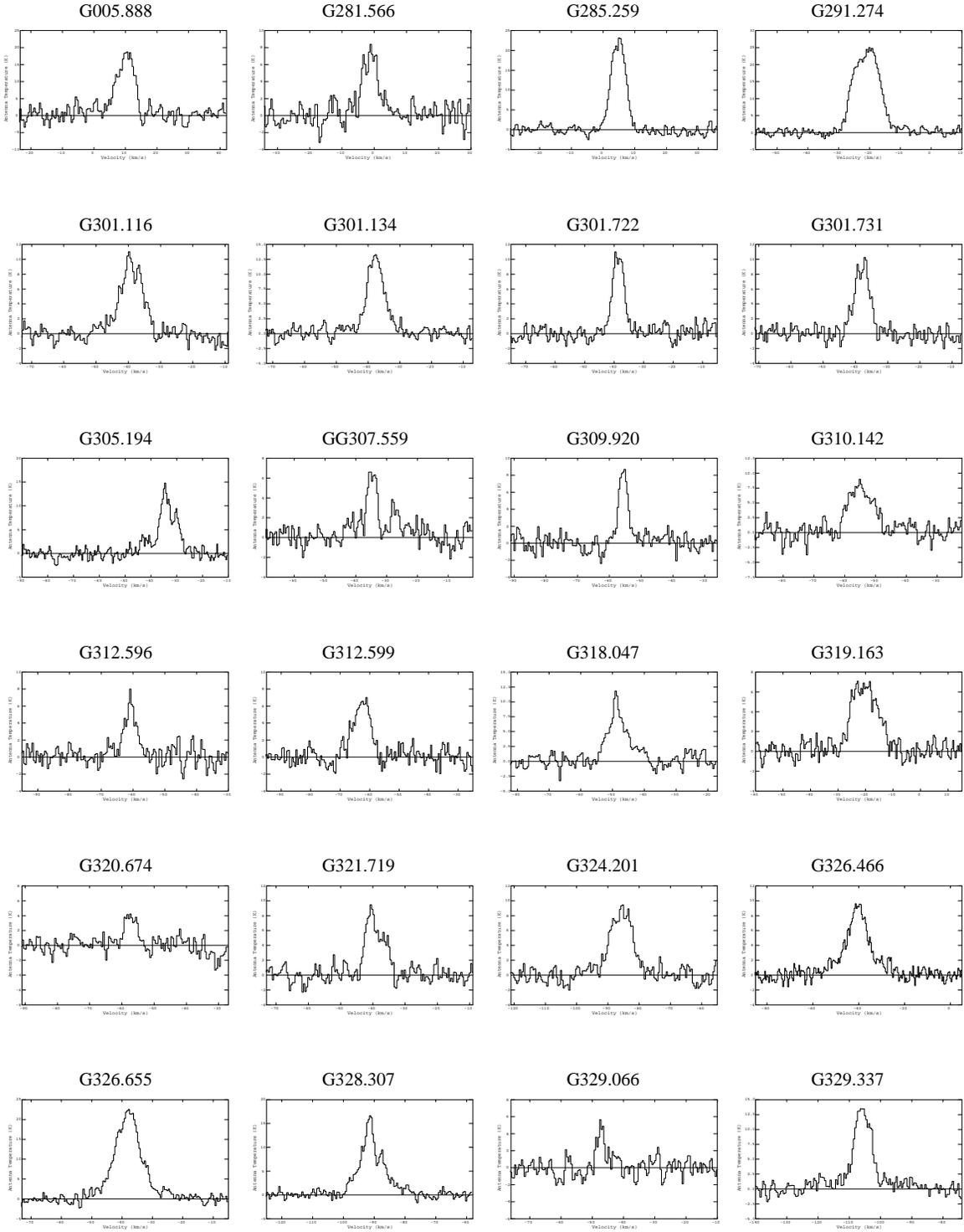
\chrisfigcaption{461.spectra.1.xfig}{Spectra toward the sources listed in Table~1.}{1a}
\chrisfigcaptioncontinued{461.spectra.2.xfig}{1b}
\chrisfigcaptioncontinued{461.spectra.3.xfig}{1c}
\chrisfigcaptioncontinued{461.spectra.4.xfig}{1c}

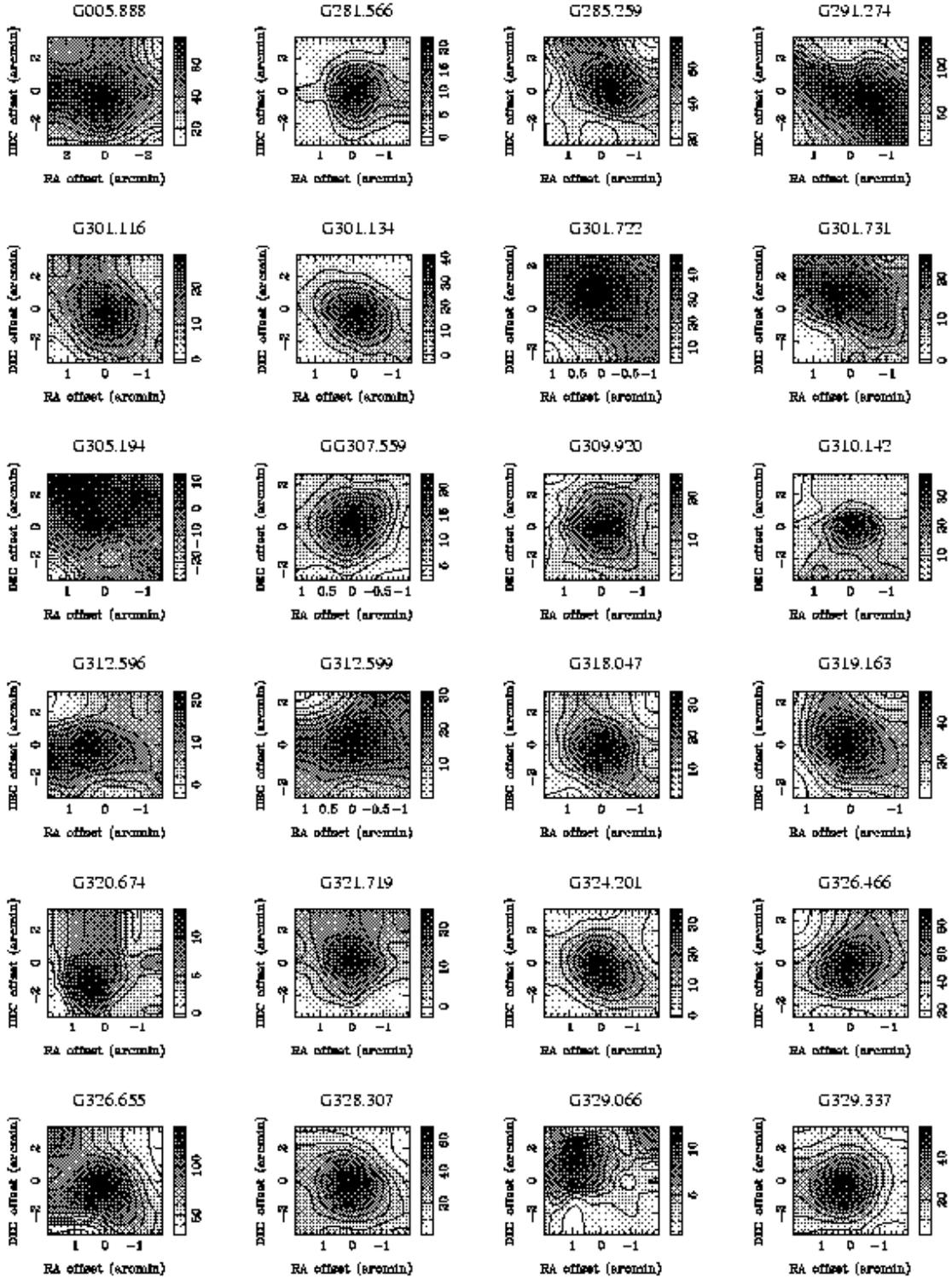
\chrisfigcaption{461.images.1.xfig}{Images of the brightest hot cores in the \co{4} line.   The figures
show the emission in the CO(4-3) line as both a greyscale and contours (5\% to 95\% of peak in 10\% steps).}{2a}
\chrisfigcaptioncontinued{461.images.2.xfig}{2b}
\chrisfigcaptioncontinued{461.images.3.xfig}{2c}

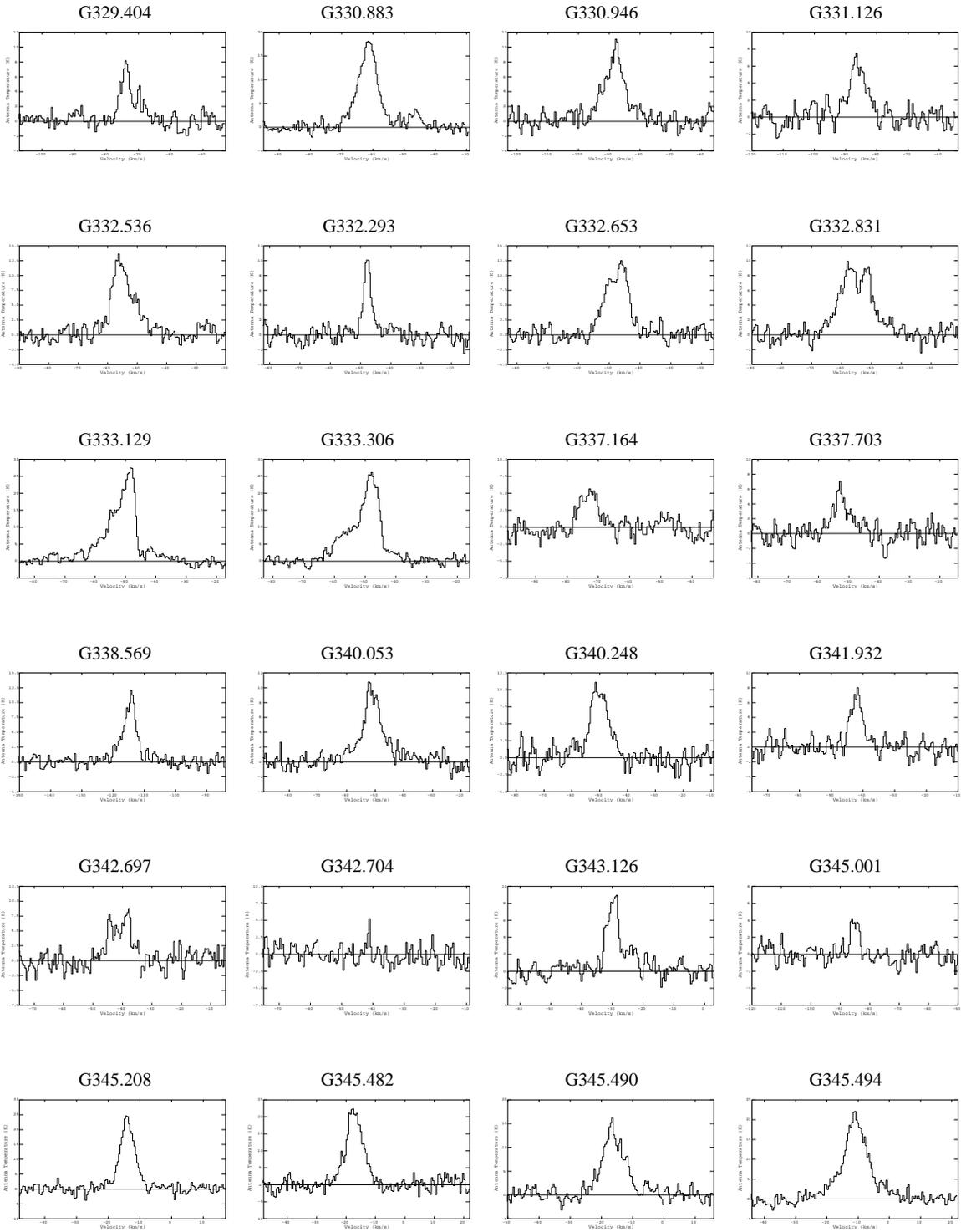
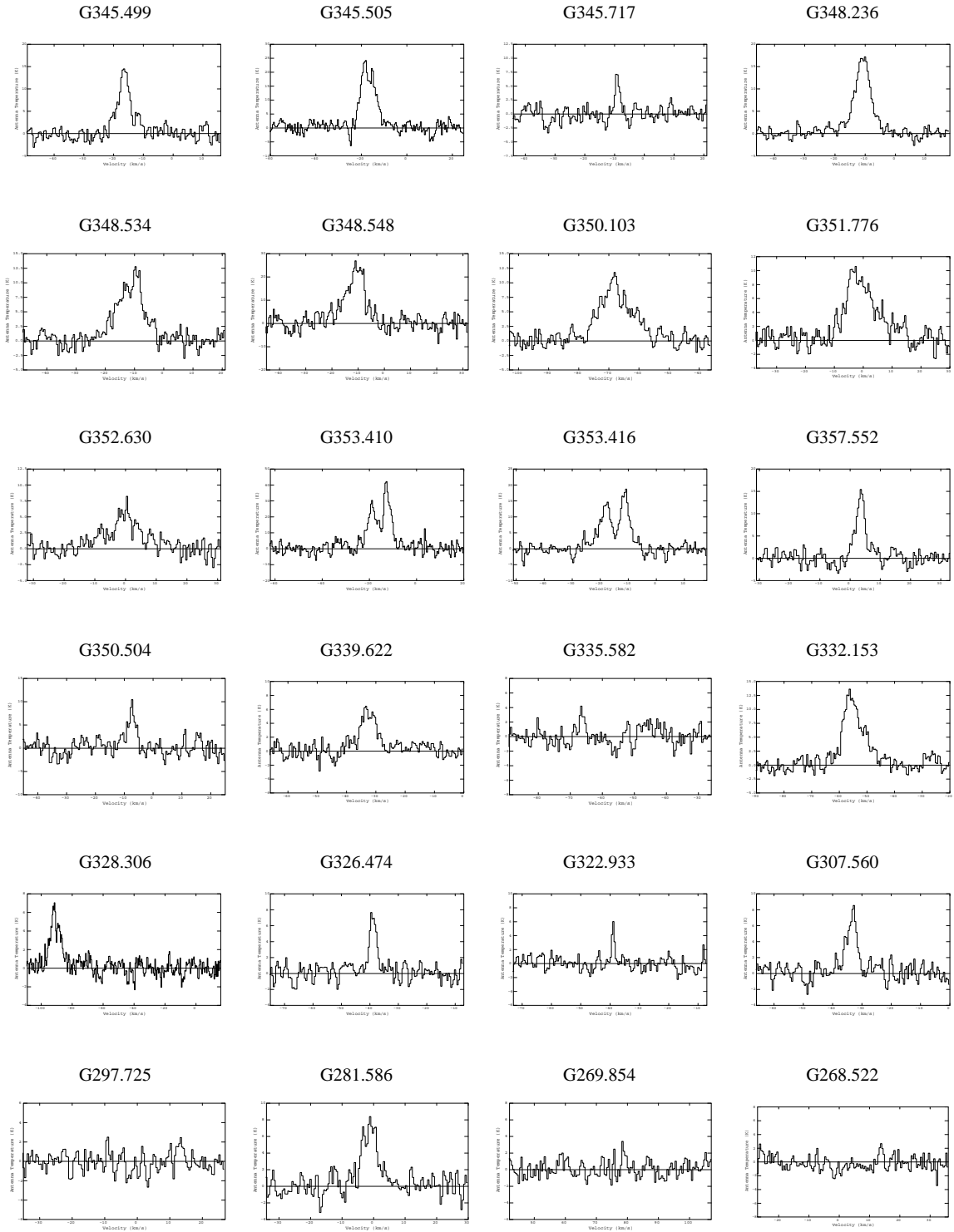
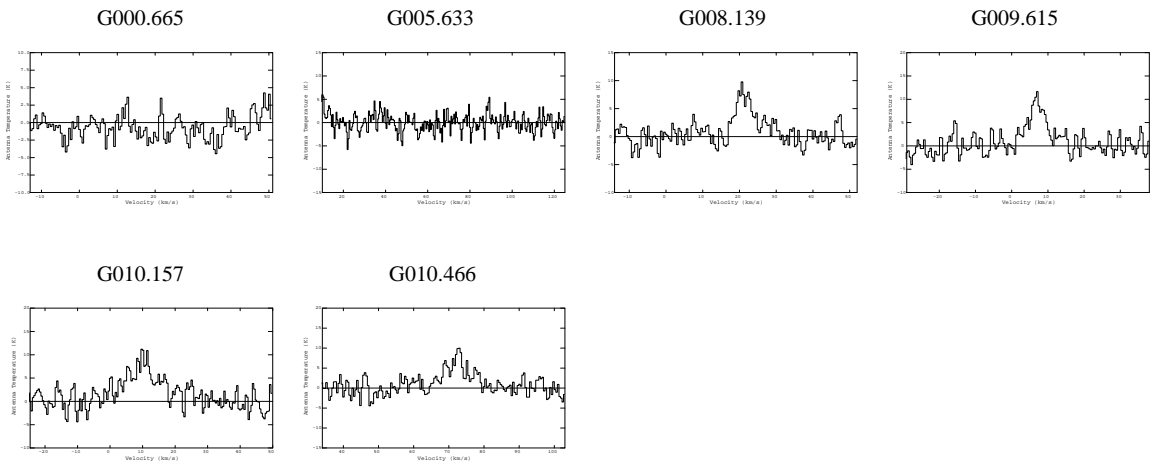
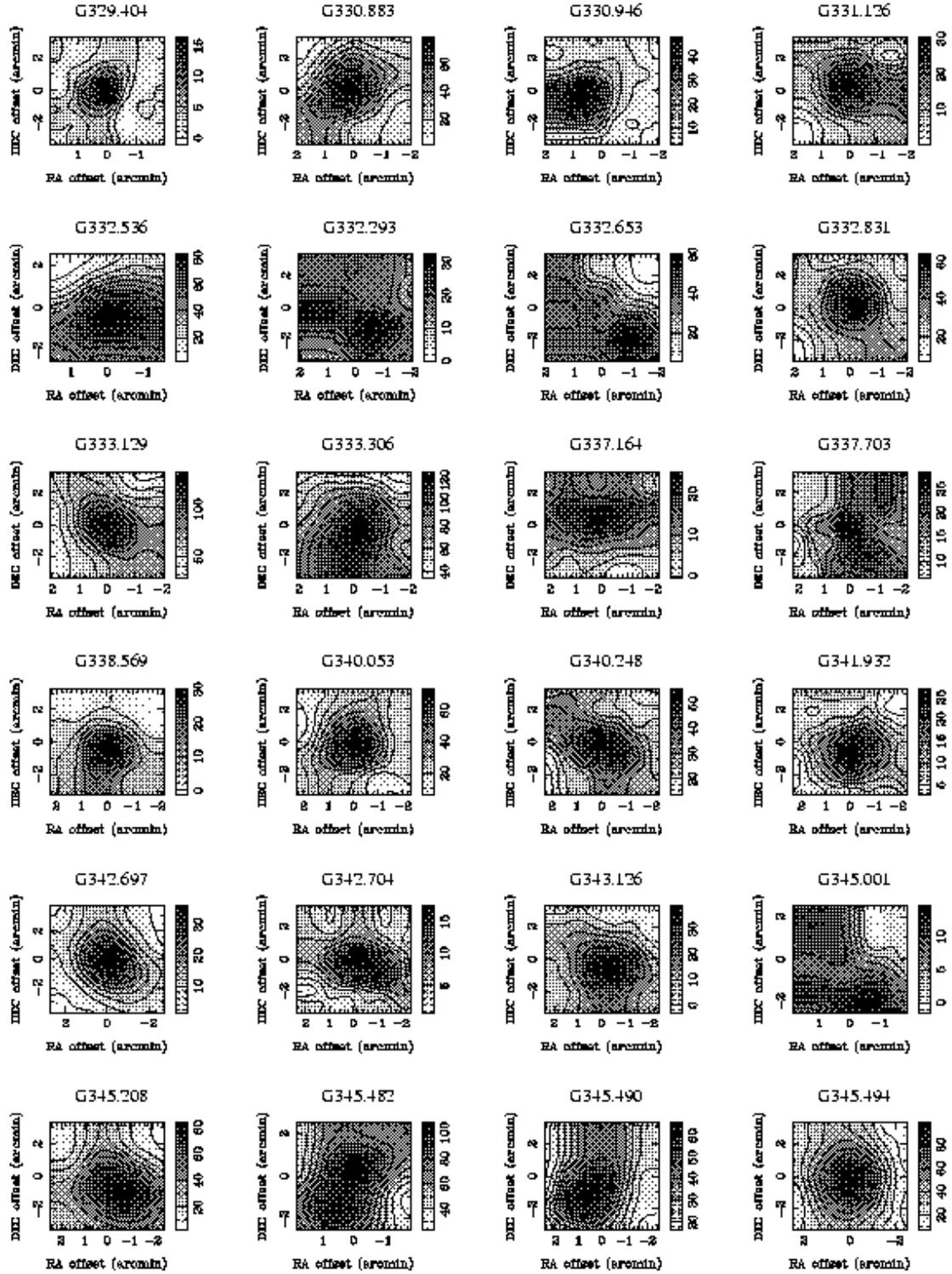
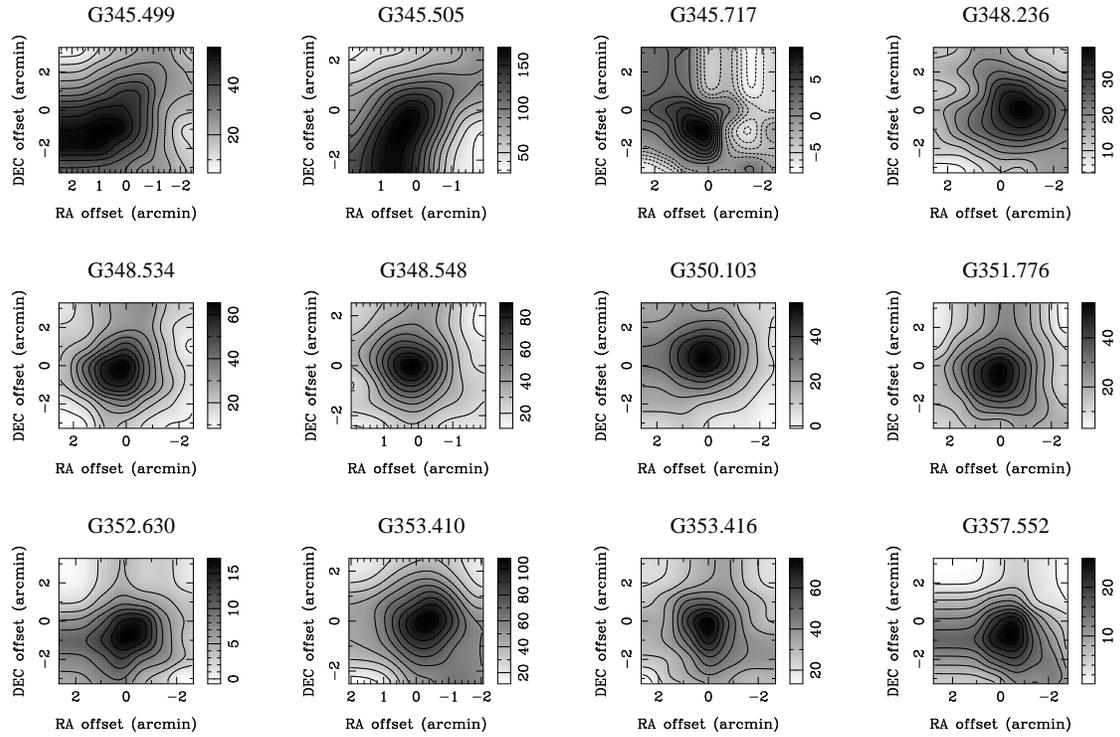
\end{document}